\documentstyle[12pt]{article}
\begin{document}
\date{}

\large

\title{Slow Magnetic Monopole: Interaction with Matter
and New Possibility of Their Detection.}
\author{I.V.~Kolokolov, P.V.~Vorob'ev\\
Budker Institute of Nuclear Physics (BINP)
\thanks{RU-630090 Novosibirsk Academician Lavrentiev Prospect 11 Russia.
P.V.VOROBYOV@inp.nsk.su}\\
V.V.~Ianovski\\
Petersburg Nuclear Physics Institute (PNPI)
\thanks{RU-188350 Gatchina St.Petersburg Russia. IANOVSKI@lnpi.spb.su}}
\maketitle

\begin{abstract}
 {\it The possibility of existence of a magnetic monopole has been
surveyed by P. Dirac even in 1931, and then from the point of view of
the modern theory by A.M. Polyakov and G.~ 'tHooft in 1974.
Numerous and
unsuccessful attempts of experimental search for  monopole  in  cosmic
rays and on accelerators in high energy particle collisions have  been
done. Also the searches have been carried out  in  mica  for  monopole
tracks as well as for relict  monopoles,  entrapped  by  ferromagnetic
inclusions in iron-ores, moon rock  and  meteorites.  These  entrapped
monopoles, when released, would have the lowest  velocities
$\beta<10^{-6}$
and do not yield ionization at all, and are hard  to  detect.
Therefore it is necessary to examine thoroughly the mechanisms of slow
heavy monopole interaction with matter and their scale of energy loss.

We discuss here the interaction of a massive  slow  magnetic  monopole
with magnetically ordered matter, with conductors, superconductors and
with condensed matter in general. Our results indicate that the energy
loss of a slow supermassive monopole reach $10^{8}~eV/cm$ and more  if
we take into consideration  the  Cherenkov  radiation  of  magnons  or
phonons and conductivity of the media. A  new  method  of  search  for
cosmic and relict monopoles by magnetically ordered film is considered
too.  This  approach  resembles  the  traditional  method  of  nuclear
emulsion chamber.  Apparently  the  proposed  method  is  particularly
attractive for detection of relict monopoles,  released  from  melting
iron ore.}
\end{abstract}

{\small PACS number(s): 14.80.H, 29.40, 77.80.D}
\vspace{0.5cm}

\section{Introduction}
\vspace{0.3cm}

A concept of a magnetic  monopole  has  been  introduced  into  modern
physics in 1931 by Paul Dirac \cite{Dir}. He postulated  existence  of
an isolated magnetic charge $g$. Using general principles  of  quantum
mechanics, he has related the electric  and  magnetic  charge  values:
$ge=\frac{n}{2} \hbar c$, where $e$ is the electron  electric  charge,
$\hbar$ is the Plank constant, $c$ is the  speed  of  light,  $n=  \pm
1,2...$  is  an  integer.  Numerous  but  unsuccessful  attempts    of
experimental search for this magnetic monopole at accelerators and  in
cosmic rays \cite{Groom, Klapdor, Barkov, D0}  have  been  done  since
then.

The new interest to this problem has arised  in  1974,  when  Polyakov
\cite{Pol} and 't~Hooft \cite{tHooft} have  shown  that  such  objects
exist as solutions in a wide class of models with spontaneously broken
symmetry. The nature of their monopoles is absolutely  different  from
the nature of other  elementary  particles,  since  they  represent  a
non-trivial topological construction of finite size, which  originates
from non-Abelian fields. So registration of the non-Dirac monopoles or
estimation of their flux limit could be an essential  contribution  to
construction of the Grand Unified Theory, and as well  it  would  give
incentives for solutions of various problems in astrophysics.

The magnetic charge of the Polyakov--'t~Hooft monopole is a  multiple
of the Dirac one $g=2 ne/\alpha$, and the value of its mass $M_{g}$
lies in the range of $10^8$ --- $10^{16}$ GeV. It is completely clear,
that the Polyakov--'t~Hooft  massive  monopoles  can  not  emerge  at
accelerators, therefore we do not  consider  accelerator  experiments.
Moreover, we assume that the monopoles that reach the surface  of  the
Earth  are  gravitationally  bound  up  either  with  the  Galaxy  (if
$\beta=v/c< 10^{-3}$) or with the Sun (when $\beta<10^{-4}$).

The monopole ionization loss in matter  has  been  evaluated  by  many
authors (look at the reviews: \cite {Groom, Klapdor, Mono}). For  fast
monopole the ionization loss appreciably (about 4700  times!)  exceeds
the loss for the minimum ionizing particles --- MIPs, which is
$dE/dl\simeq 2~MeV/g$.

In units of the ionizing loss of particle with charge e, the monopole
ionization loss is given by:
\begin{equation}
\left(\frac{dE}{dl}\right)_g
=\left (\frac{dE}{dl}\right)_{e}
\left (\frac{g}{e} \right)^2 \beta^2 ~.
\label{lost}
\end{equation}

If we  recollect  that  ionization  loss  of  a  charged  particle  is
proportional to $1/ \beta^2 $, then it is clear, that the  loss  of  a
monopole does not depend on velocity. It should be  pointed  out  here
that in GUT the monopole is a  very  heavy  particle. Therefore
supermassive monopole is practically always non-relativistic.

When $\beta \sim 10^{-3}$ the ionization loss of a monopole  decreases
to level of energy loss of MIP, at $\beta<10^{-4}$  the  ionization
mechanism of the energy loss is switched off practically,  because  in
this case the energy of monopole-atom collision already is not  enough
for ionization of the latter.

To estimate the maximum of monopole velocity $v$,  it  is  natural  to
take the velocity of the Sun relatively to the background radiation
\begin{equation}
\beta = \frac {v} {c} \simeq 10^{-3}~.
\label{vc}
\end{equation}
We shall remind here, that the virial velocity for our  Galaxy  does
not exceed $10^{-3}c$ too.

Some expansion of the ionizing detector sensitivity for slow monopoles
is reached due to the Drell effect \cite{Drell}. The  essence  of  the
Drell effect is in following. Let us assume, that a monopole passes by
a helium atom. In a strong magnetic field of  the  monopole  occurs  a
crossing of main and excited levels, due  to  Zeeman  effect.  In  the
varying magnetic field, due to non-adiabaticity of the process, it  is
possible that an electron transits to  an  excited  level.  After  the
monopole passes, some of helium atoms remain in  excited  states  with
energies of about 20  eV.  An  admixture  of  a  gas  with  ionization
potential lower than the excitation energy of helium atom gets ionized
by collision with excited helium (quenching gases - $CH_4,~ CO_2$~  or
$n-pentane$) --- the Penning effect. Therefore the  ionization  can be
detected either by direct methods, or by the radiative  recombination.
The gas counters for cosmic monopole detection on the  base  of  Drell
effect  were  created  and  used  in  a    number    of    experiments
\cite{Klapdor}. The ionization  and  Drell  effect  are  discussed  in
detail in \cite{Groom}. There is  also  a  review  of  experiments  on
search of massive monopoles.

For detection of a slow monopole  with  efficiency  close  to  1,  the
superconducting induction detectors were designed and constructed. The
pioneer in this field was B.~Cabrera \cite{Cabr}.  The  single  event,
registered  by  the  Cabrera  detector,  has  originated  a  burst  of
experimental activity of searches  for  monopoles  of  cosmic  origin.
However, the significant progress in  sensitivity  (and  corresponding
limits on the monopole flux)  has  been  achieved  only  for  ionizing
detectors \cite{Groom,MACRO}. Sensitivity of these detectors is  close
to the Parker limit \cite{Adams}
\begin{center}
$\cal F \leq$ $1 \cdot 10^{-16}(m/10^{17}~GeV)~cm^{-2} s^{-1}~sr^{-1}.$
\end{center}

The results, obtained recently by the induction experiments, are  more
modest.  At  the  flux  of  $10^{-15}~cm^{-2}sr^{-1}s^{-1}$, the
observation of only one magnetic monopole per year would  require  the
effective  area  of  a  detector  of  about  $1000~m^2$.  The   modern
superconducting  inductive  detectors  with  superconducting   quantum
interference devices (SQUID)  and  using  magnetometer  techniques  to
register the particle have an effective area  only  of  the  order  of
$1~m^2$.  Moreover,  there  are  attractive  features  of  traditional
detectors of slow moving monopoles $\beta<10^{-4}$ since normally  the
surface of composite cryogenic induction detectors is only few  square
meters.

Another  possibility  of detection of a slow GUT-monopole is the search
for proton decay induced by a heavy monopole \cite{Rub}, \cite{Cal}.
Recently  the  group working with the Baikal lake Cherenkov detector
\cite{Belo} has set the following limit on the flux of heavy magnetic
monopoles  and  the Q-balls, which are able to induce the proton decay
\begin{center}
$\cal F$ $< 3.9 \cdot 10^{-16} cm^{-2} s^{-1} sr^{-1}.$
\end{center}

In this paper we discuss the interaction of slow  moving  supermassive
magnetic monopole with ferromagnetic, conducting and other media,  and
some new features of this process. Actually we consider a  possibility
of building-up of a new type of detector of slow monopoles.  Our  idea
is based on registration of interaction of a  slow  cosmic-ray-related
monopole with the film of easy-axis and high coercitivity ferromagnet
\cite{WS}.
As a sensitive element of such a detector one can use an advanced high
density storage media, namely the magneto-optical disk  (MO  disk)
\cite{MO}. (To our knowledge  for modern MO disks an areal  density
of $45~Gbit/in^2$ has been demonstrated using  near-field  techniques,
with a theoretical possibility in excess of $100~Gbit/in^2$). The slow
monopole which is passing through the magnetic coat of MO disk, leaves
a distinctive magnetic track, and this track can be  detected  by  the
standard  polarimetric  equipment.  It  is  important  to  note   that
considerable surface can be covered by such  MO  disks.  They  can  be
exposed any reasonable time without any maintenance, like in  emulsion
chamber experiments or the CR39 nuclear  track  subdetector  of  MACRO
\cite{CR39}.

Apparently, the use of  such  passive  detectors  will  be  especially
effective  in  search  of  the  relict   monopoles,    entrapped    in
ferromagnetic  inclusions  of  iron  ore.  Such  monopoles  should  be
extracted from the ore during the melting process. These monopoles are
extracted at relatively small cross-section of the furnace and  freely
fall downwards. Such slow  moving  monopoles  can  be  detected  by  a
passive detector with MO disks. These disks are to be placed e.g. in a
cavity under the furnace. The effective exposition time, normalized to
the mass of ore, from which the monopoles are extracted, can  be  very
large. We must note, that during the exposition no detector service is
required. After exposition the MO disks should further be placed  into
specialized device to find the traces of the magnetic monopole.

Let us consider now  the  main  mechanisms  of  energy  loss  of  slow
monopole in matter.

\section{Cherenkov energy loss}

The  Cherenkov  loss  in  matter  was  considered  by  many    authors
\cite{Frank}-\cite{Zrel}. The charged  particle  energy  loss  due  to
Cherenkov radiation per unit length is
\begin{equation}
\frac{dE}{dl}=\frac{e^2}{c^2}\int\limits_{v>c_{med}}\left
(1-\frac{c_{med}^2}{v^2}\right)\omega d\omega~,
\label{cre2}
\end{equation}
where $e$ is the particle charge, $c_{med}(\omega)$ is
the velocity of electro-magnetic wave  with the frequency
$\omega$ in a matter.

For $v>>c_{med}$  Eq.(\ref{cre2}) gives
\begin{equation}
\frac{dE}{dl}=\frac{2e^2} {c^2}\omega^2 = \frac {4\pi e^2}{\lambda^2}~,
\label{cre3}
\end{equation}
here $\lambda$ is the length of a Cherenkov  radiation  wave
 in vacuum. The Cherenkov radiation of charge $e$  in
isotropic ferromagnet (ferrite) is given by
\begin{equation}
\frac{dE}{dl}=\frac{e^2}{c^2} \int\limits_{v>c_{med}} \mu \left (1-\frac
{1} {\beta^2\epsilon \mu}\right ) \omega d\omega ~.
\label{cre4}
\end{equation}

Here $\mu(\omega)$ and $\epsilon(\omega)$ are
electrical and magnetic permeabilities
of medium, $c_{med}(\omega)=c/\sqrt{\mu(\omega) \epsilon(\omega)}$

What should be expected in case of monopole  moving in  a  magnetic
medium? If
\begin{equation}
{\bf \nabla \cdot B}=4\pi\mu\rho_g~,
\label{BgKir}
\end{equation}
then for the radiation by the magnetic charge we  obtain the expression
\cite{Frank,Zrel}
\begin{equation}
\frac{dE}{dl}=\frac{g^2 \mu \epsilon} {c^2} \int\limits_{v>c_{med}}
\mu \left (1-\frac {1} {\beta^2\mu\epsilon} \right ) \omega
d\omega= \frac{g^2} {c_{med}^2} \int \limits_{v>c_{med}}\mu
\left (1-\frac {c_{med}^2} {v^2} \right ) \omega d\omega ~.
\label{crmK}
\end{equation}

Hence, for the ratio of magnetic $E_g$ to electric $E_e$ radiation we have
\begin{equation}
\frac {E_g} {E_e}=\frac {g}{e} \frac {c^2} {c_{med}^2}
\simeq 4700\cdot \frac{c^2}{c_{med}^2} ~.
\label{devK}
\end{equation}

Such an approach meets the idea of duality
$${E \leftrightarrow B}$$
and  we shall adhere just such an agreement herein.

A heavy slow monopole can not emit  usual  Cherenkov  radiation  in  a
ferromagnet, because of high phase velocity of electro-magnetic  waves
in ferromagnetic, about $c/10$, which is always much faster  than  the
monopole. However, photons are not the only particles,  which  can  be
radiated during the monopole movement in media. Besides  the  photons,
there are magnons (spin waves) in magnetically ordered media, and  the
phonons  (sound  waves)  in  any  condensed  matter.  Now  we    shall
demonstrate that the dispersion law of these excitations is such  that
they can be radiated like Cherenkov photons. And  the  coupling  of  a
magnon (phonon) to a monopole is not small and  hence  this  Cherenkov
loss is big enough.

\section {Excitation of spin wave Cherenkov radiation
by the heavy magnetic monopole }

It is well known, that a slowly  moving  heavy  monopole  cannot  emit
usual Cherenkov radiation in ferromagnetic media,  because  the  phase
speed of electromagnetic waves is of the order $c/10$  and  is  always
much faster than the monopole speed.

We shall consider the slow monopole movement in  an  ordered  magnetic
matter \cite{VK}. In such case the main mechanism  of  kinetic  energy
loss is the Cherenkov radiation of magnons. This is because the magnon
phase velocity reaches zero and the coupling of monopole to magnons is
linear in the magnon field (or in creation and annihilation operators)
and large.

For definiteness, we shall consider a ferromagnet,  but  the  estimations
below are of more general character.

The magnon's Hamiltonian in presence of magnetic  field  of  a  moving
monopole can be written in the form

\begin{equation}
H = \sum_ { \bf k } \hbar \omega_{ \bf k} a^{\dagger}_{\bf k}
a_{\bf k} + \sum_ {\bf k} \left (f_{\bf k} e^{ -i\Omega_{\bf k} t}
a^ {\dagger}_{ \bf k} + c.c \right )~,
\label {sw1}
\end{equation}
where $a^{\dagger}_{\bf k}$ is the operator of  a  magnon  creation
with a wave vector ${\bf k}$, $\omega_{\bf k}$ is its  dispersion
law, $\Omega_{\bf k}={\bf kv}$, ${\bf v}$ is the vector  of
monopole velocity and $f_{\bf k}$ is the coupling  factor of the
monopole magnetic field ${\bf B}=g{\bf \nabla}\frac {1} {r}$  to the
magnon.

The magnon energy, radiated in a unit of time, is
\begin{equation}
\epsilon =\frac {2\pi}{\hbar}\sum_{\bf k} \omega_{\bf k}{\left| f_{\bf k}
\right |}^2 \delta (\Omega_{\bf k}-\omega_{ \bf k} )~.
\label {sw2}
\end{equation}

Let the monopole velocity ${\bf v}$ be directed along the  spontaneous
magnetization,  along  z-axis.  The  general  case  is    investigated
absolutely similarly and the results will not differ much. Then
\begin{equation}
f_{\bf k}=\frac {4\pi g\mu_B} {a^{3 /
2} \sqrt {V}} \sqrt {\frac {S} {2}} \frac {k_x-ik_y } {k^2}~,
\label {sw3}
\end{equation}
here $a$ is the lattice constant, $V$ is the sample volume, $S$ is the
spin size on the node and  $\mu_B$ is the Bohr magneton.

Taking into consideration Eq.(\ref{sw3}) the equation for
$\epsilon$ can be written as

\begin{equation}
\epsilon = \frac{2 g^2 \mu_B^2 S} {a^3 \hbar} \int d^3 {\bf k} \omega_
{\bf k} \frac {k_x^2 + k_y^2} {k^4} \delta (k_z v-\omega_{\bf k})~.
\label {sw4}
\end{equation}
The integration in Eq.(\ref{sw4}) is performed over the first Brillouin
zone.

If $v\ge u$, where $u$ is the  magnon  velocity  near  the  border  of
Brillouin zone, then the magnons with large $\bf k$ are essential. Then
\begin{equation}
\epsilon \simeq \frac {\bar {\omega} g^2 \omega_M} {v}~,
\label {sw5}
\end{equation}
where the frequency $\omega_M = \frac{ 4\pi\mu_B^2 S }{\hbar a^3}$
characterizes the magnetization of media \cite{Gur},

\begin{equation}
\bar {\omega} =\frac {1} {2\pi} \int \frac {d^2\bf {k_{\bot}}}
{k_{\bot}^2 }\omega_{k_{\bot}}~,
\label {sw6}
\end{equation}
here $\bf {k_{\bot}} = (k_x, k_y)$, and $\bar{\omega}$ is close to
the  maximal frequency of magnons.

For $g^2 \simeq 4700 \cdot e^2$ we obtain

\begin{equation}
\epsilon \simeq 10^3 \cdot Ry \cdot \omega_M \bar {\omega} \tau ~,
\label {sw7}
\end{equation}
where $\tau=a/v$ is the characteristic time of interaction,
and $Ry$ is the Rydberg constant.

The typical values for magneto-ordered dielectrics are such:
$\bar {\omega } \simeq 10^{-13}s^{-1 }$, $\omega_M \simeq 10^{-11}
s^{-1}$ and for $v/c \simeq 10^{-4}$ we have $\epsilon \simeq 10^{14}
$eV/s, that makes the loss per unit of length:
\begin{equation}
\frac {dE} {dl} \simeq 10^{8}~eV/cm
\label {ocenk}
\end{equation}

From Eq.(\ref{sw5}) it is clear, that the energy loss $\epsilon$ and
$dE/dl$ grow with slowing down of the monopole. When the speed $v$
becomes $v<u$, the main contribution to the loss is done by the magnons
from "the bottom" of the spectrum.
For them, $\omega_{\bf k}=\omega_{ex} (ak)^2$, where $\omega_{ex}$  is
the    frequency,    characterizing    the    exchange     interaction
\cite{Gur,Ahi}~, and the expressions for loss become:
\begin{equation}
\epsilon=g^2 \frac{\omega_M v} {4 \omega_{ex}a^2}~;
\label{sw8}
\end{equation}

\begin{equation}
\frac {dE} {dl} = \frac {\epsilon} {v} = g^2\frac {\omega_M}
 {4\omega_{ex}a^2}~.
\label {sw9}
\end{equation}

As one can see, the energy loss per unit of length has approached a
 constant with decrease of the monopole velocity.
The characteristic velocity values will be:
$\omega_M / \omega_{ex} \simeq 10^{-2}$, $a \simeq 10^{-8}~cm~$
and for $v/c \simeq 10^{-4}$ we have again the estimation (\ref{ocenk}).

We'd like to stress, that the square-law for a magnon dispersion leads
to a non-trivial spatial structure of the Cherenkov radiation field of
the spin waves. The structure of the radiation field is similar  to  a
shock wave, but there is no radiation in  front  of  the  charge.  The
radiation field for the square-law dispersion overtakes the charge and
is non-zero in front of the charge.  This  is  because  the  quadratic
dispersion makes the group velocity of the wave faster than the  phase
velocity and faster than the velocity of the charge.

From these estimations it is clear, that the level of energy loss of a
slow  magnetic  monopole  in  magnetically  ordered  matter  can    be
comparable  to  the  ionization  loss  of  a  fast   monopole.    This
circumstance opens new opportunities for detection of  slow  monopoles
in  the  velocity  region  $v/c<10^{-4}$.  Let  us  note,  that  the
conversion mechanism of spin waves to electromagnetic ones  \cite{Ahi}
permits to detect a monopole  passing  through  a  magnetic  layer  by
traditional techniques.

\section{Excitation of Cherenkov acoustic radiation
by the magnetic monopole}

For estimation of energy loss by radiation of sound waves  (excitation
of phonons) by the monopole moving through isotropic matter, we  shall
write the Hamiltonian of an elastic system in  an  external  field  as
follows
\begin{equation}
H=\sum_{n} \frac {{\bf p}_{n}^2} {2M} +
\frac{A}{2} \sum_{n,\Delta} ({\bf x}_{n}-{\bf x}_{n + \Delta} )^2 +
\sum_{n} {\bf F}_{n} (t) {\bf x}_{n}~.
\label {aw1}
\end{equation}

Here $n + \Delta$ numbers  the closest neighbors to the node $n$,
$M$ is the mass of an ion in a lattice site, $A$ is an elastic constant
and
\begin{equation}
{\bf F}_{n} (t) = {\bf F} ({\bf r}_{n} - {\bf v} t)~.
\label {aw2}
\end{equation}

We shall estimate the strength of the force ${\bf F}  ({\bf  r}_{n})$,
acting from the monopole to the given  node,  as  follows.  First,  we
shall assume that this force is located on one node (due to its  short
range):
\begin{equation}
{\bf F (r_{n})} = {\bf F} \delta_{n0}~.
\label {aw3}
\end{equation}

Secondly, at rest this force causes deformation, and the affected node
is shifted by
\begin{equation}
\delta a \sim \frac {F} {A}~,
\label {aw4}
\end{equation}
and, assuming the deformation energy to be
$\epsilon_{def} \sim A\delta a^2,$
we have the force as
\begin{equation}
{\bf F} \sim A\sqrt { \frac{\epsilon_{def}} {A}} \sim
\sqrt{\epsilon_{def} A}~.
\label {aw5}
\end{equation}

The Hamiltonian Eq.(\ref{aw1}) can be  expressed  similar  to
Eq.(\ref{sw1})

\begin{equation}
H = \sum_{\bf k} \hbar \omega_{\bf k} a^{\dagger}_{\bf k}a_{\bf k} +
\sum_{\bf k} \left (f_{\bf k} e^{-i\Omega_{\bf k} t}
a^{\dagger }_{\bf k} + c.c \right ) ~,
\label{aw6}
\end{equation}
but now:
$a^{\dagger}_{\bf k}$ is the operator of phonon creation with the wave
vector ${\bf k}$, $\omega_{\bf k} $ is its dispersion,
$\Omega_{\bf k}={\bf  kv}$,  ${\bf  v}$  is  the  vector  of  monopole
velocity, and $f_{\bf k}$ is  the  coupling  factor  of  the  magnetic
monopole field ${\bf B}=g\nabla \frac {1}{r}$  to  the  phonon.  We
shall write the following expressions for them:

\begin{equation}
f_{\bf k} =
\frac{1} {2i} \hbar^{1/2} \frac {1} {\sqrt {N}}\cdot \frac{F}
{( D_{\bf k} M)^{1/4}}~,
\label {aw7}
\end{equation}

\begin{equation}
\omega_ {\bf k } = \sqrt {\frac{2 D_{\bf k}}{M}}~,
\label{aw8}
\end{equation}

\begin{equation}
D_{\bf k} = \frac{A} {2} \sum_{\bf {\Delta}} \left | 1-e^{i{\bf k
\Delta }} \right | ~.
\label{aw9}
\end{equation}
Here $N$ is the number of lattice sites in the crystal.

Accordingly, the energy of phonons, radiated in one unit of  time,  is
equal to
\begin{equation}
\epsilon = \frac {a^3} {4 (2\pi)^2} \int {d^3 {\bf k} \frac {F^2}
{(D_{k} M)^{1/2}}\omega_{\bf k} \delta (k_z v-\omega_ {\bf k})}~,
\label{aw10}
\end{equation}
where $a$ is the lattice constant $a^3=V/N$, $V$ is the
sample volume.

For the essentially supersonic monopoles, integration over $dk_z$
gives the factor $1/v$, and the Eq.(\ref{aw10}) results to:
\begin{equation}
\epsilon = \frac {a^3} {4 (2\pi)^2 } \frac {1} {v} \int {d^2{\bf k}_{\bot}
\frac{F^2} {(D_{{\bf k}_{\bot}} M)^{1/2}}
\omega_{{\bf k}_{\bot}}} =
\frac{a} {2\sqrt{2} v} \frac{F^2} {M}~.
\label {aw11}
\end{equation}

Using the evaluation of Eq.(\ref{aw5}), for $F$ we shall obtain:
\begin{equation}
\epsilon \simeq \epsilon_{def} \frac {a} {v} \frac {A}{M}~.
\label {aw12}
\end{equation}

Now from decomposition of Eq.(\ref{aw9}) at small ${\bf k} $ and using
Eq.(\ref{aw8}) it is possible to express $A/M$ through  the  speed  of
sound. As a result Eq.(\ref{aw12} ) acquires the form
\begin{equation}
\epsilon \simeq \epsilon_{def} \frac {1} {Z} \frac { c_{s}} {v}
\frac{c_{s}} {a} \simeq \epsilon_{def} \frac{ c_{s}} {v} {\bar
{\omega}_a}~,
\label {aw13}
\end{equation}
where ${\bar {\omega}_a}$ is the frequency limit for phonons,
$Z$ is the number of nearest neighbors.

If $\epsilon_{def} \sim Ry$, $c_{s}/v \sim 0.1$ and
${\bar {\omega}_a} \sim 10^{13}~s^{-1}$
is about the Debye temperature in energy units, then:
$$\epsilon \simeq 10^{13} eV/s,$$
$$\frac{dE}{dl}\simeq 10^{7} eV/cm,$$
which is a little less than the loss by radiation of magnons.\\

\section{Interaction of a massive monopole with a superconductor}

The monopole leaves "a tail"-  a  string  of  magnetic  field,  as  it
traverses a superconductor. It is the  Abrikosov  vortex,  which  is
filled by the normal phase and by  the  monopole  magnetic  flux.  The
vortex core radius is determined by the coherent length $\xi$, and the
effective  radius  of  magnetic  flux  tube,  which  is  the    London
penetration depth $\lambda_L$
\begin{equation}
\lambda_L=\sqrt {\frac{mc^2} {\mu n_s e^2}}~.
\label{rLond}
\end{equation}

Thus, the energy loss of a slow monopole in a superconductor is
made up of two components: \\
- moderation by magnetic string tension, \\
- loss due to destruction of the condensate of Cooper pairs \\
  and formation of the normal-phase core.

The magnetic string loss is
\begin{equation}
\frac{dE_m} {dl} = \frac{1} {8\pi} \frac{\Phi_0^2}
{\pi \lambda_L^2}~.
\label{mlos}
\end{equation}

Using a typical value of $\lambda_L = 10^{-5}cm$, and
$\Phi_0 =4\cdot 10^{-7}~ G\cdot cm^2$ we obtain
\begin{equation}
\frac{dE_m} {dl} \simeq 10^{-5}~erg/cm \simeq 10^7~eV/cm ~.
\label{mlos1}
\end{equation}

The loss due to formation of the core of normal phase is
\begin{equation}
\frac{dE_n}{dl} = \Delta E n_s \pi \xi_c^2~,
\label{nlos}
\end{equation}
here  $\Delta  E$  is  the  gap  width,  $n_s$  is  the  density    of
superconducting carriers.  For  typical  suprconductor  parameters  we
estimate the  value  of  the  energy  loss  to  be  of  the  order  of
$10^7~eV/cm$.

To be accurate, it is necessary to include the surface energy  on  the
core border of the Abrikosov vortex into $\frac{dE_n}{dl}$. However it
is clear, that the  result  will  remain  within  the  same  order  of
magnitude.

At the London length $\lambda_L = 10^{-5}~cm$  the  pressure  of  the
magnetic field is relatively small, and the  loss  for  the  Cherenkov
acoustic radiation can be neglected.

\section{Monopole interaction with conductors}

When the monopole transverses a normal massive conductor,  finally  it
loses energy on behalf of the Joule heat. Besides that, the moderating
force acts on it in the same manner as in a superconductor,  from  the
magnetic field of the induced currents.

The thickness of a skin layer in  a  conductor  is  described  by  the
expression
\begin{equation}
\delta = \sqrt {\frac{c^2\tau} {2\pi\sigma\mu}}~,
\label{skin}
\end{equation}
where $\sigma$ is the conductivity, $\mu$ the permeability, and $\tau$
is the period (the pulse duration) of the magnetic field.

We can express $\tau$ as
\begin{equation}
\tau=\frac{\delta}{v}~,
\label{tau}
\end{equation}
where $v$ is the monopole velocity.
Substituting Eq.(\ref{tau}) into Eq.(\ref{skin}) we obtain the expression
\begin{equation}
\delta_c = \frac{c^2} {2\pi \sigma \mu v }~.
\label{skinc}
\end{equation}

This length has a simple physical meaning -- at  distances  less  than
$\delta_c$, the monopole field can be considered to be free,  with  an
accuracy to the permeability of the matter, but at  the  distances  of
$\delta_c$ and more, it will form the magnetic tail of  the  monopole,
like the string in a superconductor. Due to the finite conductivity of
the matter, the tail gradually dissipates, and energy of the  magnetic
field transforms into Joule heat.

It is easy  to  show,  that  for  a  typical  conductor,  the  maximal
current-density of a metal layer with thickness  $\delta_c$  is  large
enough to keep the magnetic flux of the monopole  in  a  tube  with  a
radius of the order $\delta_c$.

With the known critical length, it is easy to  evaluate  the  monopole
energy loss in a conductor. Within the cross section of  the  diameter
equal to the critical length, the magnetic field is
\begin{equation}
B_c = \frac {\Phi_0} {\pi\delta_c^2 }~.
\label{Bcsk}
\end{equation}

Accordingly, the monopole  energy  loss  due  to  the  tail  formation
(friction in the magnetic field ) is
\begin{equation}
\frac{dE_m} {dl} = \frac {\Phi_0^2} {8\pi^2\delta_c^2 } =
\frac{\Phi_0^2\sigma^2\mu^2\ v^2} {2 c^4}~.
\label{Wmsk}
\end{equation}

Examining Eq.(\ref{Wmsk}), we can see that the monopole energy  loss
in the conductor is proportional to the square of the monopole velocity.
At $v=10^{-4}c$ the loss is
\begin{equation}
\frac{dE_m} {dl} \simeq 10^{-5}~erg/cm \simeq 10^7~eV/cm~.
\label{mloskin}
\end{equation}

\section{Track formation by the slowly moving monopole in
a ferromagnetic}

It is expected that a slow monopole,  moving  transversely  through  a
magnetized ferromagnetic film, should leave  a  distinctive  track  of
magnetization in it. We can use this phenomenon to design an effective
detector of supermassive cosmic monopoles. For this purpose  we  shall
consider now in detail the mechanism of magnetic  track  formation  by
such a type of monopoles.

\subsection{Quasistatic approximation.}

Let  us  consider  a  thin  layer  of  easy-axis
\footnote{Let us recall that the term "easy-axis" means that the
anisotropy energy of a magnet has the
form $\epsilon_a=-K\sum_n \left(S_n^{z}\right)^2$, where the easy axis
is $z$-axis, $S_n^z$ is $z$--component of the ion spin placed on the lattice
site $n$ and the anisotropy constant $K$ is positive. The spontaneous
magnetization of such a  magnet is oriented along the easy axis.}
hard   ferromagnetic
magnetized perpendicularly to the surface along an easy  axis.  It  is
easy to see  that  the  external  magnetic  field  is  absent  (double
layer!), but the surface density of a  magnetostatic  energy  of  such
configuration is rather large.

Therefore such a magnetization configuration  appears
to have too high energy, and the magnetic  film  is  split  into
a  system  of
magnetic domains. Magnetizations of the domains are also orthogonal to
the surface and are directed along the easy axis,  and  have  opposite
signs in neighboring domains \cite{Tika}.

The domain characteristic size is determined by a condition of minimal
total energy per unit of the film surface.  For    films  of ferromagnetic
garnets a
characteristic scale of domain structure is about  1/100~mm.  And  the
total energy of the  film  decreases  more  than  1000  times  due  to
creation of the domain structure.

However, if the anisotropy constant $K_u$ is reasonably large, and the
effective field of anisotropy exceeds the value of demagnetizing field
\begin{equation}
\frac{2 K_u} {I_s} \ge \frac {I_s} {\mu_0}~,
 ~ ~ ~ ~ K_u \ge \frac {1} {2} \frac{I_s^2} {\mu_0}~,
\label{metstab}
\end{equation}
then the system  is  in  a  metastable  state.  Therefore  for  domain
formation,  it  is  necessary  to  have  a  magnetic    bubble    with
magnetization opposite to the film magnetization and parallel  to  the
demagnetizing field. In  Eq.(\ref{metstab})  the  parameter  $I_s$  is
magnetization of the film material and $\mu_0$ is permeability.

It turned out, that the size of such magnetic bubble should be  finite
and not very small. For simplicity we consider a  magnetic  bubble  of
cylindrical shape of radius $r$ and with axis orthogonal to  the  film
surface and its magnetization opposite to the film magnetization.

Let us introduce the characteristic length $r_0$:
\begin{equation}
r_0 = 2\mu_0 \gamma / I_s^2
\label{r00}
\end{equation}
Then the total energy of the magnetic bubble can be found as
\begin{equation}
{\cal E} = 2\pi\gamma rh\left((1-2 N) \frac{r}{r_0} -1\right),
\label{esum}
\end{equation}
where $N$ is a demagnetization factor of the domain, $r$ is the radius
of the domain, $h$ is the film thickness, $\gamma$ is the  density  of
surface energy of the domain wall.

The magnetic bubble size can be evaluated  from the condition
\begin{equation}
\frac{\partial {\cal E}} {\partial r}=0~.
\label{var}
\end{equation}
If we use for the demagnetization factor the following
approximation:
\begin{equation}
N=\frac{2}{3}\frac{r}{h}~,
\label{N}
\end{equation}
then we find  two  values  $r$  of  the  domain  radius  as  solutions
Eq.(\ref{var}).

The first value is  the radius of collapse  $r_c$, below which the
domain-magnetic bubble is unstable and collapses
\begin{equation}
r_c=
\frac{h} {4} \cdot \left [1-\left(1- \frac {4 r_0} {h}
\right)^{1/2}\right]~.
\label{rcoll}
\end{equation}
And the second solution gives us  the equilibrium radius  $r_{eq}$
\begin{equation}
r_{eq} = \frac {h} {4} \cdot \left [1+\left (1-\frac{4 r_0}{h}
\right)^{1/2}\right]~.
\label{req}
\end{equation}
At $h >> r_0$ we can find from (\ref{rcoll}) and (\ref{req})
\begin{equation}
 r_{c}|_{h > r_0} \rightarrow \frac{r_0} {2}~;
\label {rch}
\end{equation}
and
\begin{equation}
r_{eq}|_{h > r_0} \rightarrow \frac {h} {2}.
\label {reqh}
\end{equation}
We see that $r_0$ defines
the  minimal  radius  of
collapse.
From  non-negativity  of  the  radicand   in(\ref{rcoll}) we obtain,
that the minimal width of a ferrolayer, in which a cylindrical domain
(further -  magnetic bubble) can exist, is
\begin{equation}
h_{min} = 4 r_0 = 8 \frac {\mu_0 \gamma} {I_s^2}~.
\label{hmin}
\end{equation}

It is necessary to emphasize, that in a zero external magnetic  field,
the magnetic bubble  is  unstable  and  turns  into  a  stripe  domain
\cite{Eshenf}.

Let us evaluate the orders of magnitudes:
$$\gamma \simeq 1~ erg/cm^2,$$
$$I_s^2 \simeq~10^6 erg/cm^3 ~.$$
So $r_0$ will be of the order of  $10^{-6}~cm$,  or  10  nm.  And  the
collapse radius is yet less for "thick" enough films!  Thus,  at  film
thickness  about  100  nm  we  shall  have  a  magnetic  bubble   with
characteristic size of the order of 30 nm.

What is the field of a monopole at such distance?
\begin{equation}
H= \frac{\Phi_0} {4\pi r_0^2} \simeq 2\cdot 10^3~Oe~,
\label{bmon}
\end{equation}
 that is enough without any doubt for re-magnetization of a
material with coercitivity of the order of 1~Oe.

Let us note, that the existence of  the  minimal  radius  hinders  the
fluctuational creation of microscopic domains, the  magnetic  bubbles.
In this sense, a homogenically magnetized film  can  be  quite  stable
against  splitting  into  a  structure  of  magnetic   domains.    All
abovementioned is true for films with high mobility of  domain  walls.
Films with low wall mobility are even more stable,  and  at  the  same
time, domains with radius less than the collapse radius can  exist  in
them, in principle. So, in the $Co/Pt$ films the  movement  of  domain
walls is suppressed. And in the 20 nm film the transverse  domains  of
cylindrical form with diameter of the order of 50-100 nm are obtained.
We wish to note, that the coercitivity of the easy-axis  $Co/Pt$  film
is of the order of 1-2~kOe \cite{M-opt}.

\subsection{Track dynamics --- domain formation}

However, our speculations in previous  subsection  are  true  only  in
static, for very slow monopoles only. As we  have  noted  before,  the
characteristic velocity of a monopole is $v \simeq  10^{-4}-10^{-3}c$,
and for our consideration let us  assume  $v=10^{-4}c$.  The  time  of
monopole interaction with an electron $\tau$ can  be  defined  as  the
time, during which a field  higher  than  some  critical  field  $H_c$
interacts with the electron
\begin{equation}
\tau \simeq \frac{r_c}{v} \simeq \frac{1} {v} \sqrt {\frac {\Phi_0}
{4\pi H_c}} ~.
\label{tmon}
\end{equation}

At $H_c$ of the order $3\cdot10^3$ Oe we have
$\tau \sim 3\cdot 10^{-12} ~s.$
It means, that the spin-flip of the magnetic in the "track" takes
place during the interaction time.

For such spin-flip, the  adiabatic  condition  is  necessary  since  the
frequency of spin precession in the magnetic field of the moving monopole
should  be  much larger than the inverse time of the interaction
\begin{equation}
\omega = \frac {\mu_B H} {\hbar} \gg \frac{1}{\tau}~.
\label{wmon}
\end{equation}

It is possible to derive from here the minimal magnetic field which is
appropriate for the adiabatic mode, and the track radius:

\begin{equation}
H \gg H_c = \frac{\hbar}{\mu_B \tau} = \frac{4\pi\hbar^2 v^2}
{\mu_b^2 \Phi_0}~,
\label{Hc}
\end{equation}

\begin{equation}
R_t \ll r_c = \sqrt {\frac{\Phi_0} {4\pi H_c}}~.
\label{rt}
\end{equation}

In our case at $v \simeq 10^{-4}c$ we get $$H_c \simeq
10^{7}Oe~;~~r_c \simeq 10^{-7}cm,$$
and for $v \simeq 10^{-6}c,$ we have
$$H_c \simeq 10^{3}Oe~;~~ r_c \simeq 10^{-5}cm.$$

It is obvious, that the conditions of adiabatic and even resonant spin
flip are not fulfilled, while $r_c < r_0$,  that  corresponds  to  the
monopole speed $v \simeq 10^{-6} c$.

We shall consider the  influence  of  the  conductivity  of  the  film
material now. The reason is that the monopole magnetic flux  is  being
frozen into the cylindrical  area  around  the  track  axis  and  then
spreads radially. The radius $\delta_c$ of the flux pipe
and the diffusion factor
of the flux are determined by the film conductivity.
The value of $\delta_c$ we can estimate as
\begin{equation}
\delta_c =\frac{c^2} {2\pi\cdot \sigma\mu v} ~.
\label{skinc1}
\end{equation}

As it was noted before, this length has a simple physical meaning.  At
distances less than $\delta_c$ the monopole field can be considered as
free. At distances of the order of $\delta_c$ and more,  the  magnetic
tail of the monopole is formed, which is an analogue of a string in  a
superconductor. Due to the finite conductivity of material,  the  tail
spreads gradually, and the energy of the magnetic field converts  into
heat.

At the monopole velocity about $v \simeq 10^{-4}c$, the flux pipe  has
the radius of the order of $10^{-5}cm$. The flux pipe blows with  time
as:

\begin{equation}
R(t) = \delta_c \sqrt {\frac{t}{\tau}}~.
\label{rfromt}
\end{equation}

Thus the magnetic moment of the track is conserved,  as  well  as  the
frozen flux. It is easy to calculate  the  average  intensity  of  the
magnetic field in the flux pipe immediately after monopole flight

\begin{equation}
H = \frac{\Phi_0} {\pi \delta_c^2} \sim 10^{3} Oe ~.
\label{Hpip}
\end{equation}

The typical time of the monopole interaction with an  electron  in
a conductor is:

\begin{equation}
\tau \simeq \frac {\delta_c}{v} \simeq 10^{-11}-10^{-10}~s~,
\label{timpip}
\end{equation}
and the field strength,  providing  the adiabatic inversion of the
magnetic spin in a track, will be:

\begin{equation}
H_c \simeq \frac {\hbar} {\mu_B \tau } \simeq 10^{4}~Oe~.
\label{Hcrit1}
\end{equation}

Thus, the frozen field in the conductor $H_c$ can  affect  appreciably
the process of spin-flip  in  the  track  and  provide  the  adiabatic
spin-flip of electrons  in  the  magnetic  at  monopole  speeds  below
$10^{-4}~c$. Besides, it can render a certain influence on dynamics of
the domain walls in conducting films with high mobility.

\subsection{Detection of the monopole track}

As it was shown in this  paper,  the  domain  induced  by  the  moving
monopole has the typical size of the order of 50~nm and  magnetization
about several thousand Gauss. Then the domain magnetic  flux  $\Phi_d$
will be of the order:

$$\Phi_d = \pi r^2\cdot I_s \simeq \Phi_0=2\cdot 10^{-7}~G\cdot cm^2$$

For detection of such a flux we can use the high  sensitive  fluxmeter
on the basis of superconducting quantum interference device such as
SQUID,
or magneto-optical device on the basis of Kerr effect (rotation of the
polarization plane of light reflected by a surface  of  a  ferromagnet
which is magnetized perpendicular to the surface). It  is  clear  that
the second way is technically easier and does not require a  cryogenic
maintenance. In the later case, the realization  of  such  a  detector
requires a surface covered with a thin layer of  easy-axis  magnetized
magnetic media, plus a magnetooptic device  to  detect  the  spots  of
transverse magnetization of  the  film  (to  detect  the  domain  with
opposite  direction  of  magnetization!)  with  a  system  of  precise
positioning.

A similar technique has emerged recently in  an  almost  ready  shape,
suitable for the detector design with minimal adjustment.  It  is  the
magneto-optic recording technology used in modern magneto-optic  disks
(MO disks) and their readout devices. Already there are MO disks  with
multilayer  coating  of  $Sm/Co$  and  $Pt/Co$.  The  coercitivity  of
multilayer coats $Pt/Co$  is  about  $1KOe$  at  10  layers  of  total
thickness  about  $15~nm$~\cite{M-opt}. The size of a magnetic bubble
which  can  be  detected  in  such  a  coat  by  the
magneto-optical method is about $60~nm$.  This  techniques  using  the
near-field optics has been designed, f.i. in Bell Laboratories \cite{http}.

The coercitivity of coats with $Sm/Co$ is in the  interval  $3-5~kOe$,
and the reference size of magnetic bubble is $50~nm$~\cite{M-opt}. For
detection of the magnetic track of the monopole it is possible to  use
slightly modified standard MO-drives, as  we  have  mentioned  before.
Having covered a large enough surface  with  such  MO  disks,  we  can
obtain an effective  and  relatively  nonexpensive  detector  of  slow
moving space monopoles, which we can expose during unlimited time.

However, it is more effective to use the  MO-detector  to  search  for
relict monopoles, entrapped in ferromagnetic inclusions of Fe ore
\cite{Detect}.
Naturally,  the  melted  iron  ore  becomes  paramagnetic   and    the
ferromagnetic trap disappears. Then  the  monopole  is  likely  to  be
surrounded by a cluster of several dozens of iron atoms. The size of a
complex is determined by thermodynamic equilibrium:
\begin{equation}
\frac{\mu_{Fe}\cdot g}{r_{Fe}^2} = \frac{3}{2} kT
\label{balance}
\end{equation}
and the radius of the iron atomic complex, paramagnetically bound to
the monopole at $T\approx1200^0~C$ is:
\begin{equation}
 r_{Fe} \simeq 6 \cdot 10^{-8} cm.
\label{diam}
\end{equation}

Apparently the complex is small and contains about 30 atoms of Fe.
Considering, that the movement of such small blob is determined by
the Stokes law:
\begin{equation}
F_v = 6\pi \cdot r_{Fe}\cdot \eta  v~,
\label{stocs}
\end{equation}
where $v$  is the monopole velocity, $\eta$ is the dynamic viscosity
of liquid foundry iron,
$\eta = 2\cdot 10^{-3} kg/(m \cdot sec)$ at $T = 1250^0~C$.

Equating the force of friction to the gravity $F_g = mg$, we find the
velocity $v$ of the monopole falling through the melt
\begin{equation}
v = \frac{mg}{6\pi \cdot r_{Fe} \cdot \eta}~,
\label{radius}
\end{equation}
that makes  $v \simeq 3\cdot 10^{-1} m/sec$  for a monopole with
mass about $10^{15}~GeV$.

This corresponds to kinetic energy of the complex about $10^6 ~eV$,
which is large  enough for a skinning off of the complex at the solid
bottom of the furnace. Let's remark, that from a formal point
such an approach is quite acceptable, as the Reynolds number
in our case it is not large enough: $Re<10^{-3}$.
This grain (complex) should sink in the liquid  iron  at  10-100  cm/s
velocity until it reaches the bottom of the blast  furnace.  Then  the
atoms of iron are stripped off the monopole in  the  material  of  the
oven bottom, and the monopole falls further, accelerating  up  to  the
velocity of sound in the matter.

Usually the blast furnace melts about 10 000 tons of Fe ore per day, and it
could be easely expose the MO-detector, for example, during one year.
Thus we hope that such MO-detector can improve significantly the
experimental limit on the density of relict monopoles entrapped in
Fe ore, which today is equal to $\rho_M < 2\cdot 10^{-7}/g$~\cite{PDG}.

Furthermore, in a sinter machine the ore is also heated above the Curie
temperature, but not up to the melting point. So, to shake off the iron
atoms, we have to kick  the iron ore pieces with acceleration of
$10-100~g.$ Clearly, in this case the probability  of monopole release is
considerably lower.

\section{Conclusion}

In this paper we analyze the mechanisms of energy loss of supermassive
slowly moving monopole in matter. In general, this study concerns  the
magneto-ordered materials, e.g. ferromagnets, as well as conductors  in
normal and in superconducting phases. The new effective  mechanism  of
energy loss of slow magnetic monopole is considered: the  Cherenkov
excitation  of  spin  waves  during  monopole  movement  through   the
magneto-ordered magnetic.

The interaction of monopoles  with  films  of  magnetic  materials  is
considered. In particular, the interaction of slow monopoles with thin
films of easy-axis magnetics with high  and  low  mobility  of  domain
walls (materials with magnetic bubbles)is discussed. It is shown, that
during the movement of a slow monopole through the magneto-hard
magnetic film, a track-domain can  be  formed with typical size of about
50~nm  and  with  magnetization  of  about
several thousand Oe. Thus the magnetic flux of the track appears to be
about the value of the flux quantum. For detection of such a flux, the
detectors using fluxmeter on the basis of already widely  known  SQUID
can be used.

However, in our opinion, for registration of  traces  of  slow  cosmic
monopoles in magnetic matter, the experimental devices using the  Kerr
magneto-optical  effect  are  more  appropriate.  They  have   emerged
recently in a shape suitable for detector design, with an  appropriate
adjustment.

Apparently, such passive detectors will  be  especially  effective  in
search of relict monopoles, entrapped in ferromagnetic  inclusions  of
iron ore. These monopoles should be  extracted  from  the  ore  during
melting process. Then these slow moving monopoles can be detected by a
passive MO detector. We can expose MO disks in a cavity under a  blast
furnace  exactly  under  the  bath  with  melting  metal,  where   the
temperature does not exceed $+50^0~C$.  In  the  melting  process  the
temperature of ore exceeds  the  Curie  point  and  its  ferromagnetic
properties disappear. Hence the ferromagnetic  traps  which  hold  the
monopoles are "switched off" and the released monopoles  fall  through
the melting metal to the bottom of the bath and finally through the MO
disks. While a monopole, moved downwards by the gravity force  crosses
the surface of one of the MO disks, it leaves a magnetic track in  its
coat. It is possible to obtain the slow moving relict monopole also by
the sinter machine.

The modern blast furnace have capacity of the order of 10 000 tons
of pig-iron per day. The arrangement of the MO-detector under such
furnace allow us to considerably improve the recent experimental
limit on the relict monopole density entrapped in Fe ore.

The authors are grateful to all  colleagues  for  helpful  and  lively
discussions of this work in various  places  and  institutes.  Special
thanks go for stimulating interest and useful remarks  to  L.M.~Barkov
and I.B.~Khriplovich (both from Budker INP), and  G.~Tarle  (Univ.  of
Michigan).


\newpage
\begin{thebibliography}{99}
\bibitem{Dir} P.A.M. ~Dirac, Proc. Roy. Soc. {\bf A133} (1931) 60.
\bibitem{Groom} D.E.~Groom, Phys. Rep. {\bf 140} (1986).
\bibitem{Klapdor}  H.V.~Klapdor-Kleingrothaus, A.~Staudt,
{\it Teilchenphysik ohne Beschleuniger}, Stuttgart, 1995
\bibitem{Barkov} L.M.~Barkov,I.I.~Gurevich, M.S.~Zolotarev et al.,
Zh. Eksp. Teor. Fiz {\bf 61} (1971) 1721, [JETP 61 (1971) 1721]
\bibitem{D0} D0 Collaboration, S.~Abachi et al., preprint Fermilab
             Pub-98/095-E,(1998).
\bibitem{Pol} A.M.~Poljakov, Pis'ma Zh. Eksp. Teor. Fiz. {\bf 20}
(1974) 430.\\ JETP Lett. 20 (1974) 194.
\bibitem{tHooft} G.~'t~Hooft, Nucl.Phys. B {\bf 79}, 276 (1974).
\bibitem{Mono} {\it Dirac monopole}. Ed. B.M.~Boltovsky  and  Yu.D.~Usachev,
Moscow,1970.
\bibitem{Drell} S.D.~Drell et al., Phys. Rev. Lett. {\bf 50} (1983) 644.
\bibitem{Cabr} B.~Cabrera, Phys. Rev. Lett. {\bf 48} (1982) 1378.\\
               R.D.~Gardner et al., Phys. Rev. {\bf 44} (1991) 622.
\bibitem{MACRO} MACRO collaboration, Phys. Lett. {\bf B406} (1997) 249.
\bibitem{Adams} F.C.~Adams et al., Phys. Rev. Lett. {\bf 70} (1993) 2511.
\bibitem{Rub} V.A.~Rubakov, Pis'ma v ZhETF {\bf 33}(1981)644,\\
              Nucl. Phys. {\bf B203} (1982) 311
\bibitem{Cal} C.~Callan, Phys. Rev. D {\bf 25} (1982) 2141.
\bibitem{Belo} I.A.~Belolaptikov et  al.  LANL  e-print  archive
astro-ph/9802223.
\bibitem{WS} I. Kolokolov, P. Vorob'ev, V. Ianovski. {\it
Interaction of a massive slow magnetic monopole with matter}. In:
Proc. of the XXXII PNPI Winter School. (Particles and Nuclear
Physics). Ed.: J. Azimiv, V. Bunakov, V. Gordeev. SPb, 1999 (134-152).
\bibitem{MO} Ed.: Glenn T. Sincerbox, James M. Zavislan,
{\it (Selected Papers on Optical Storage)}, Vol. MS 49, SPIE - The
International Society for Optical Engineering, 1992.
\bibitem{CR39} B. Ichinose et al. Nucl. Inst. Meth.{\bf A286}(1989)327.\\
MACRO Collaboration, S. Ahlen et al.,~ Search for magnetic  monopoles
with MACRO track etch detector",~ LNGS 94/115 (1994).
\bibitem{Frank}I.M.~Frank, {\it Vavilov---Cherenkov
radiation},Moscow, 1988.
\bibitem{Kirzh} D.D.~Kirchniz, V.V.~Losjakov, Pis'ma Zh. Eksp. Teor.
Fiz. {\bf 42} (1985), 226.
\bibitem{Zrel} V.P. ~Zrelov, {\it Vavilov---Cherenkov radiation and its
 application to high energy physics}, Moscow, 1968.
\bibitem{VK} P.V.~Vorob'ev, I.V. ~Kolokolov. {\it Cherenkov radiation
of magnon and phonon by the slow magnetic monopole}.
Preprint BINP 98-16, Novosibirsk 1998; LANL  e-print  archive hep-ph/9806495,\\
Pis'ma Zh. Eksp. Teor. Fiz. {\bf 67} (1998) 866.
\bibitem{Gur} A.G. ~Gurevich, G.A. ~Melkov,
{\it Magnetic oscillation and waves}, Bica Ratton: CRC Press, 1996.
\bibitem{Ahi} A.I. ~Ahiezer, V.G. ~Bar'jakhtar, S.V.~Peletminski,
{\it Spin waves}, Moscow,1967.
\bibitem{Tika} S.~Tikadsumi,~{\it Physics of ferromagnetism}, Kashili Kaisha
Su\"{o}kab\"{o}, Tokio, 1984.
\bibitem{Eshenf} A.~Eschenfelder,~{\it  Magnetic bubble technology},
Springer-Verlag Berlin, Heidelberg, NY. 1981.
\bibitem{M-opt} Chung-Hee Chang, M.H.Kryder.
J.Appl. Phys. {\bf 75} (1994) 142.
\bibitem{http} http:$//
portal.research.bell-labs.com/\\leisure/souvenirs/gallery/bits.html$
\bibitem{Detect} V.V.~Ianovski, I.V.~Kolokolov, P.V.~Vorob'ev.
{\it Detection of Slow Magnetic Monopole.}
Preprint PNPI EP-38-1999, N2323, Gatchina 1999;\\
LANL  e-print  archive hep-ph/9909528.
\bibitem{PDG} Review of Particle Physics. Evro.Phys.J. 3 (1998) 742,
PRD 54 (1996) 686.
\end{thebibliography}
\end{document}